\documentstyle[osa,manuscript]{revtex}
\begin{document}

\title{Electric field gradients in
$s$-, $p$- and $d$-metal diborides and the effect of pressure 
on the band structure and T$_c$ in MgB$_2$ }

\author{
N.I. Medvedeva$^{1,2}$, A.L. Ivanovskii$^1$,
J.E. Medvedeva$^{2,3}$, A.J.Freeman$^2$, D.L. Novikov$^4$}

\address{
$^1$Institute of Solid State Chemistry, Ekaterinburg, Russia}

\address{
$^2$Department of Physics and Astronomy, Northwestern University,
Evanston, Illinois}

\address{
$^3$Institute of Metal Physics,  Ekaterinburg, Russia} 

\address{
$^4$Arthur D. Little, Inc. Cambridge, Massachusetts, 02140}

\maketitle
\begin{abstract}
Results of FLMTO-GGA (full-potential linear muffin-tin orbital --
generalized gradient approximation) calculations of the band
structure and boron electric field gradients (EFG) for the new
medium-T$_c$ superconductor (MTSC), MgB$_2$, and related diborides
MB$_2$, M=Be, Al, Sc, Ti, V, Cr, Mo and Ta are reported. The
boron EFG variations  are found to be related to
specific features of their band structure and particularly to the
M-B hybridization. 
The strong charge anisotropy at the B site 
in MgB$_2$ is completely defined by the 
valence electrons - a property which sets MgB$_2$ apart from other diborides.
The boron EFG in MgB$_2$ is weakly dependent of applied pressure:
the  B p electron anisotropy increases with pressure, 
but it is partly compensated by the increase of  core charge assymetry. 
The concentration of holes in bonding $\sigma$ bands
is found to decrease slightly  from 0.067 to 0.062 holes/B 
under a  pressure of 10 GPa.
Despite a small decrease of N(E$_F$), 
the Hopfield parameter increases with pressure 
and we believe that  the main reason for the
reduction under pressure of the superconducting 
transition temperature, T$_c$,
is the strong pressure
dependence of  phonon frequencies, which is sufficient to
compensate the electronic effects. 
\end{abstract}

%PASC numbers: 74.72.-h

\newpage

Recently, Akimitsu et al \cite{c1} reported the discovery of
medium-T$_c$ superconductivity (MTSC) with T$_c$ of about 39 K in
magnesium diboride (MgB$_2$) with a simple composition and crystal
structure (AlB$_2$-type, space group P6/mmm, Z=1). 
The band structure
calculations showed \cite{c3,c4,c5,c6,c7,c8} that the MTSC in
MgB$_2$ can be attributed to a strong electron-phonon coupling, a
rather high density of states from 2D (in-plane) metallic boron
$\sigma (p_{x,y})$ bands  at E$_F$ and the existence of
$p_{x,y}$-band holes.
By now, a number of studies have been performed with NMR
\cite{c9,c10,c11} which is a very powerful technique for
investigating  the properties of  MgB$_2$. The 
measured quadrupole interaction is determined by  the value of
the electric field gradient (EFG), which is directly related to
the  charge distribution  around the  nucleus. 
Thus, theoretical EFG studies are  important in order
to give a reliable interpretation of the experimental data
based on the electronic structure.

The  EFG at the boron site in MgB$_2$ was  found 
 experimentally to be  much larger
than those  for $d$- diborides (Table I).
As seen from Table I, the variation in EFG for diborides 
covers two  orders of  magnitude 
and shows  trends
which cannot be explained by the crystal structure changes.     
For example, these EFG's in 
AlB$_2$ and TiB$_2$ 
differ  by  almost three times, 
but their lattice parameters  $a$ and $c$ are approximately the same;
and, vice versa, equal EFG's were obtained in BeB$_2$ and MgB$_2$,
for which the lattice parameters have  
the largest differences among diborides
under consideration. 
No attempts was made previosuly to relate the EFG changes 
to the peculiarities of electronic structure.

In this paper, we present  results of first-principles
full-potential LMTO-GGA \cite{c12} calculations (within the
generalized gradient approximation (GGA) for the exchange correlation
potential) of the electronic structure and  EFG's at the
boron site for  MgB$_2$ and other $s$-,$p$-,$d$-diborides.
We compare the calculated EFG's  with the experimental 
and other theoretical data and 
explain the EFG's variation  based on the anisotropy of 
boron 2p occupancies.
The pressure dependence of the EFG, 
which is very sensitive to the charge distribution, represents a good test
for the anisotropy study.    
The experimental and theoretical data on the
electronic and elastic behavior of MgB$_2$ under pressure 
are contradictory: it was found to be nearly isotropic \cite{Loa,c24},
anisotropic \cite{Islam} or strong anisotropic \cite{Rabin,c23}. 
We simulated the pressure effect  on the band structure of MgB$_2$  
and estimated  the changes in 
the EFG, boron p-occupancies, hole concentration 
and Hopfield parameter under pressure. 
These investigations are of interest, 
since the pressure dependence of the T$_c$ is the key difference 
between
the conventional BCS \cite{c4} and hole \cite{Hirsh} 
superconductivity mechanisms.
  
The band structures of MgB$_2$ and some other diborides are shown
in Fig.1.
There are two distinct sets of B 2$p$-bands: $\sigma$ (2$p_{x,y}$)
and $\pi$ ($p_z$)-types with considerably different dispersions.
The  B $2p_{x,y}$ bands are quasi-two dimensional (2D) along the
$\Gamma$-A line in the Brillouin zone (BZ) 
and  make a considerable contribution to the density of states 
at the Fermi level, N(E$_F$),
for MgB$_2$. Now was shown \cite{c4,c5,c6,c7,c8}, that the existence
of degenerate $p_{x,y}$-states above E$_F$ at the $\Gamma$ point
in the BZ is crucial for the MTSC in diborides. 
The high Tc is explained by the strong coupling of these
holes to the in-plane E$_{2g}$ phonon modes \cite{c7,kong}. 
The B
$2p_z$-bands are responsible for the weaker $pp_\pi$-interactions
and these 3D-like bands have maximum dispersion along $\Gamma-A$.
The bonding and antibonding B $p_z$ bands cross E$_F$ at the K
point
and their location and dispersion depend on the M-B hybridization. For
BeB$_2$ (Fig.1b) and AlB$_2$ (Fig.1c), the $p_{x,y}$ bands  are,
respectively, partly and completely filled, the Fermi surface
topology changes and medium -T$_c$ superconductivity is absent
\cite{c13,c14}.
The $p_z$-bands progressively move down in going from BeB$_2$ to
MgB$_2$ and AlB$_2$, demonstrating the strengthening of M-B
bonding.

The band structure and chemical bonding of all
3$d$, 4$d$ and 5$d$-metal diborides were previously investigated in
detail \cite{c16,c17,c17a,c17b}. These studies showed that the cohesive
properties of AlB$_2$-type diborides are explained in terms of the
band filling. The Fermi level for TiB$_2$ (ZrB$_2$ and HfB$_2$)
falls in the pseudogap where bonding states are occupied and
antibonding states are empty (Fig.1d). The shift of  E$_F$
results in the partial emptying of the bonding states (ScB$_2$,
Fig.1e) or the occupation of antibonding states (VB$_2$, CrB$_2$,
MoB$_2$, TaB$_2$). Both cases correspond to the lowering of the
cohesive properties (melting temperature, enthalpies of formation,
etc.). From first-principles estimates of M-M, M-B and B-B
bonding strengths, we found that the cohesive energy of
$d$-diborides with filled bonding states 
decreases when the atomic number increases in the row
due to the weakening of M-B hybridization
and that the AlB$_2$-type diborides of the group VII and VIII elements are
unstable\cite{c16,c17,c17a,c17b}. Thus, the main feature of the
band structure of TM diborides is the progressive filling of B
$p_{x,y}$, $p_{z}$-bands and the dominant role of $d$-states near
E$_F$.

A systematic search for superconductivity in the $d$-diborides
(with M = Ti, Zr, Hf, V, Ta, Cr, Mo)
showed that T$_c$  is
below $\sim 0.4$ K \cite{c18}. Only NbB$_2$ and ScB$_2$ were found
to be superconductors with a T$_c$ of about 0.6 K \cite{c18} and
1.5 K \cite{c15}, respectively. Recently, a relatively high critical
temperature, T$_c$ $\sim$ 9 K, was found in TaB$_2$ \cite{kacz}.
As seen in Fig.1f, the bonding states in TaB$_2$ are fully
occupied, the Fermi level is shifted away from the pseudogap to the
region of antibonding states and  Ta 5$d$ states define  
N(E$_F$) (0.9 states/eV). Among the 3$d$-diborides, only
in ScB$_2$ are the 2$p_{x,y}$ bands  not completely filled (there
is a small hole concentration of these states at $A$), but they
lie below E$_F$ at $\Gamma$ and the largest contribution to N(E$_F$)
arises from Sc 3$d$ states. Based on the band structure results and
calculations of the electron-phonon interaction \cite{kong}, one
may conclude that   superconductivity with medium T$_c$ is
unlikely in undoped diborides, except for MgB$_2$; the absence of
holes in the two-dimensional $\sigma$-bands at $\Gamma$ results in 
hardening of zone-phonon modes and weak electron-phonon coupling.

The electric field gradient tensor, defined as 
the second derivative of the electrostatic potential at the nucleus,
was calculated directly from 
the  FLMTO charge  density.
The calculated principal components of the boron  EFG tensor, 
$V_{zz}^B$, 
are shown together with other theoretical and
experimental data in Table 1. Note that the asymmetry parameter, 
$(|V_{xx}|-|V_{yy}|)/|V_{zz}|$, is equal to 0 for the AlB$_2$-type
structure. 

The largest boron EFG's in MgB$_2$ and BeB$_2$
demonstrate the strongest assymetry of the charge distribution
as compared with other diborides
(note, that here and below we consider the absolute value of EFG's). 
For the $3d$-diborides, the
EFG decreases from ScB$_2$ to TiB$_2$ and increases when going
from TiB$_2$ to VB$_2$ and CrB$_2$. 
The boron EFG's for 4$d$ (MoB$_2$) and 5$d$ (TaB$_2$) diborides are
much smaller than the EFG's for isoelectronic 3$d$ diborides. 
Note that all calculated EFG's are in
very good agreement with available experimental data and with FLAPW
\cite{Wimmer} theoretical results \cite{c19} (Table 1).

To analyze the variation  of the boron EFG's in diborides, 
we consider it as a sum of
electron ($V_{zz}^{el}$) and lattice (ion) ($V_{zz}^{lat}$)
contributions (Table 1).
For  the $s,p$ diborides,
the  ion contribution is small,
and the boron EFG is mainly determined by
the anisotropy of the valence electrons. 
For the $d-$diborides, $V_{zz}^{lat}$
is larger  and  depends somewhat on the  metal,
(except ScB$_2$ where the lattice parameters are the largest).
For TaB$_2$, the electron and ion contributions 
are almost equal,  the EFG is positive and         
smallest among all diborides.
The electronic contributions explain  the boron EFG variation,
although they overestimate the calculated and observed EFG's;
clearly, the lattice contributions must be taken into account 
in order to obtain  good agreement with experiment for $d$ diborides.
Among the diborides, the lattice contribution is the smallest for
MgB$_2$ and  
the  strong charge anisotropy  in MgB$_2$
is completely defined by the valence B $p$ electrons --
a property which sets MgB$_2$ apart from other diborides.  

A qualitative explanation of EFG behavior 
may be given based on the  anisotropy of B 2$p$ partial 
occupancies, $\Delta n_p$=$p_z-(p_x+p_y)/2$,
since 
$V_{zz}^{el}$ $\sim$  $<1/r^3>$ $\Delta n_p$
and one may consider the boron $p$ 
$<1/r^3>$ expectation value  to be constant for all diborides
discussed. 
Thus, the variation of the electronic EFG's is determined by the
interplay of $p_z$ and $p_x, p_y$ occupations.
As seen from the partial DOS (PDOS)
obtained by means of a Mulliken population analysis
, the $p_{x,y}$ orbitals are
more occupied than are  $p_z$ orbitals (Fig.2) and 
$V_{zz}^{el}$ is negative for all diborides considered. For MgB$_2$ and
BeB$_2$, the high  $p_{x,y}$  peaks lying below -2 eV
lead to large negative $\Delta n_p$ values and, therefore, to  large boron
EFG's. The small increase of $p_z$ occupancy explains the EFG lowering
for MgB$_2$ (and AlB$_2$) as compared to BeB$_2$. Thus, the weaker
M-B bonds for $s$-, $p$-diborides correspond to  larger 
boron EFG's.

For $d$-diborides, the $p_z$ PDOS is more localized due to 
strong covalent M 3$d$-B 2$p$ bonding and 
the  high peak  at 3-5 eV below E$_F$ 
decreases $\Delta n_p$ and $V_{zz}^{el}$ compared with $s$-, $p$-diborides. 
Among the $3d$-diborides, the EFG is  smallest for
TiB$_2$, which has the strongest p-d hybridization. 
Weaker p-d hybridization 
(less intense p$_z$ peak) 
for ScB$_2$
and CrB$_2$ results in a larger anisotropy,  $\Delta n_p$,
and the boron EFG's are larger for these compounds than for  TiB$_2$.
Since 4$d$ and 5$d$ states are less localized
than are 3$d$ states, the corresponding  B $p_z$ PDOS are broadened for
MoB$_2$ and especially for TaB$_2$ (Fig.2)
(the strong hybridization of B 2p and Ta 5d states was shown also in
Ref.\cite{Pick}), that leads to the small EFG's. 
While a Mulliken analysis is not an accurate approach for
the calculation of orbital charges, especially for delocalized
p-orbitals, it still allows one to describe general trends in EFG's and to
correlate them with  peculiarities of  the electronic structure.
Thus, we conclude  that  M-B$p$
hybridization is the main factor controlling the boron EFG variation.

The effect of hydrostatic pressure on the EFG at the B site in MgB$_2$ 
was investigated for 5 and 10 GPa with
lattice parameters taken from  the extrapolation formula
\cite{c23} 
$a=a_0(1-0.00187P)$ and $c=c_0(1-0.00307P)$.
We found a very slow increase of EFG with pressure
(Table 1),
that also demonstrates that the boron EFG's 
in diborides do not have a strong dependence on
the interatomic distances. 
As  the EFG is a very sensitive characteristic, 
no large changes are expected in the anisotropy of the B charge distribution 
 under pressure.

As expected, the boron $p$  bands  widen under pressure (Fig.1a).
One can see that the band shifts relative to E$_F$ 
are different for different directions of BZ;
the main changes in the occupied $p_{x,y}$ and
$p_z$ bands occur in the low energy range 
at the M, K and A points. 
These bands  
move down with pressure relative to E$_F$ along  $\Gamma$-M-K-$\Gamma$
and  A-L  and 
the overall shift of the PDOS to lower energies 
leads to the loss of these states in the energy range from E$_F$ to -2 eV, 
as stated  in Ref. \cite{c24}. 
The decrease of $p_{x,y}$ and  $p_z$  PDOS 
near  E$_F$ is
partly compensated by its increase at lower energies,
and as a result, the changes in the partial $p$
occupations are small. 
We found the increase of  $p$ occupancies with pressure
to be  anisotropic -- the larger growth of $p_{x,y}$ occupancy 
compared to $p_z$  giving  
an $\Delta n_p$ increase by 0.02 for 10 GPa. 
The accurate calculation
gives smaller EFG than it follows from $\Delta n_p$. 
As seen from the Table,
the  increase of B p electron anisotropy
is partly compensated by the increase of  the core charge contribution,
and as a result
the boron EFG in MgB$_2$ is weakly dependent on applied pressure.
Thus, we  conclude, that 
the  charge distribution at B site 
shows more  isotropic change under pressure
than do the B $p$ valence electrons. 
due to the compensating behavior of electron and core systems.
NMR measurements under pressure would be important
to confirm  this conclusion.
 
The PDOS changes  near E$_F$ under a pressure of 10 GPa is shown in Fig.3. 
The hole concentration in the bonding $\sigma$ bands 
decreases by 0.005 (within an energy interval up to 0.8 eV).
The changes in $p_{x,y}$ PDOS  is due to  
the behavior of these bands with pressure along 
$\Gamma$-A:
the bands move down at $\Gamma$ and up at A, 
resulting, respectively,  in the loss of holes 
in the energy range up to 0.5 eV
and their increase for higher energies.
Although  the hole $p_z$  PDOS change 
is negative in the energy interval up to 0.2 eV,
its concentration is almost constant under pressure 
for energies  up to 0.8 eV.  
The loss of both $p_{x,y}$ and $p_z$ states near E$_F$
results in a small decrease of N(E$_F$) by 0.02 states/eV.
Thus, we showed that the carrier concentration  decreases 
and so the observed decrease of the resistance with pressure 
is likely  to be connected  
with better coupling between the sintered grains, 
as  suggested in Ref. \cite{Mont}.

Finally, we estimated the pressure dependence of
the Hopfield parameter, $\eta$, which is an electronic part of the
electron-phonon coupling
$\lambda$ = $\eta$/M$<\omega^2>$, where
$\eta$= N(E$_F$)$<I^2>$.   
The calculation of 
the averaged electron-ion matrix element squared, $<I^2>$, 
performed  within the rigid muffin tin
approximation\cite{foot} gave a faster increase of $<I^2>$ with pressure 
than the N(E$_F$) decrease.
As a result, 
despite a small decrease of N(E$_F$), ($dN(E_F)/dP$ = -0.51\%/GPa), 
the Hopfield parameter increases with pressure as
$d\eta/dP$ = + 0.55\%/Gpa. 
Hence the decrease  of N(E$_F$) cannot be considered 
as the reason for  the T$_c$ reduction,
which is known \cite{chu} to behave 
as $dT_c/dP$ = - 1.6 K/GPa. 
Thus,
according to the McMillan formula, the main reason for the
reduction of T$_c$ under pressure is the strong pressure
dependence of  phonon frequencies, which is sufficient to
compensate for the electronic effects.

Work at Northwestern University supported by the U.S. Department
of Energy (grant No. DE-F602-88ER45372)

 \begin{figure}
 \caption{Band structures of (a) MgB$_2$ (dot lines for zero pressure, 
  solid lines for 10 GPa),
  (b) BeB$_2$, (c) AlB$_2$, (d) TiB$_2$, (e) ScB$_2$,
  and (f) TaB$_2$. \label{fig1}}
 \end{figure}

\begin{figure}
\caption{
Boron partial densities of $p_{x,y}$ (dash lines) and $p_z$
  (dot lines) states (PDOS) and their anisotropy (solid lines) for
   (a)  MgB$_2$, (b) BeB$_2$, (c) TiB$_2$, (d) TiB$_2$. 
 The Fermi level corresponds to the zero energy. \label{fig2}}
 \end{figure}
 
\begin{figure}
\caption{
The change in the boron $p_{x,y}$ (solid  line) and $p_z$ (dash line)
  PDOS in Mg$B_2$ under  pressure 10 GPa. 
   The Fermi level corresponds to  the zero energy. \label{fig3}}
\end{figure}

\begin{table}
\caption{
Theoretical and experimental boron EFG, $V_{zz}$ (in
$10^{21}V/m^2$),
  for $s$-, $p$-, $d$-diborides}
\begin{tabular}{cccccc}
Diboride   & $V_{zz}^{el}$ &  $V_{zz}^{lat}$ & $V_{zz}$ & $|V_{zz}^B|$\cite{c19} & $|V_{zz}^B|$, exp \\ \hline
MgB$_2$     & -1.94 & 0.06 &  -1.88 &  &  1.69\cite{c9}  \\
MgB$_2$\footnotemark[1] & -2.00 & 0.10 & -1.90  &  &   \\
BeB$_2$     & -2.43 & 0.33 & -2.10 &  &            \\
AlB$_2$     & -1.17 & 0.18 & -0.99 &   &  1.08\cite{c21}  \\
ScB$_2$     & -0.75 & 0.13 & -0.60 &   &           \\
TiB$_2$     & -0.66 & 0.31 & -0.35 & 0.38   &  0.37\cite{c22}  \\
VB$_2$      & -0.76 & 0.38 & -0.38 & 0.39   &  0.43\cite{c22}  \\
CrB$_2$     & -1.01 & 0.42 & -0.59 & 0.60   &  0.63\cite{c22}  \\
MoB$_2$     & -0.55 & 0.32 & -0.23 & 0.22   &  0.23\cite{c22}  \\
TaB$_2$     & -0.21 & 0.25 &  0.04 & $<$0.05  &  0.02\cite{c22}  \\ 
\end{tabular}
\footnotetext[1]{under pressure 10 GPa}
\end{table}


\begin{references}
\bibitem{foot} For zero pressure we obtained the Hopfield parameter
 $\eta$ = 129.2 mRy/$a^2_B$ which
 is very close to value 135 mRy/$a^2_B$ from \cite{c4}
 and leads to approximately the same
 T$_c$ $\sim$ 20 K for the Coulomb pseudopotential $\mu$=0.1
 and phonon frequency $<\omega_{log}>$=500 cm$^{-1}$.
\end{references}
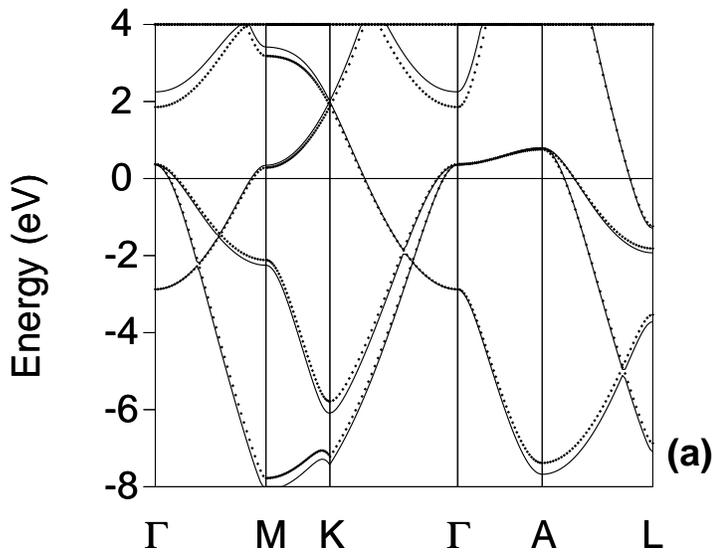
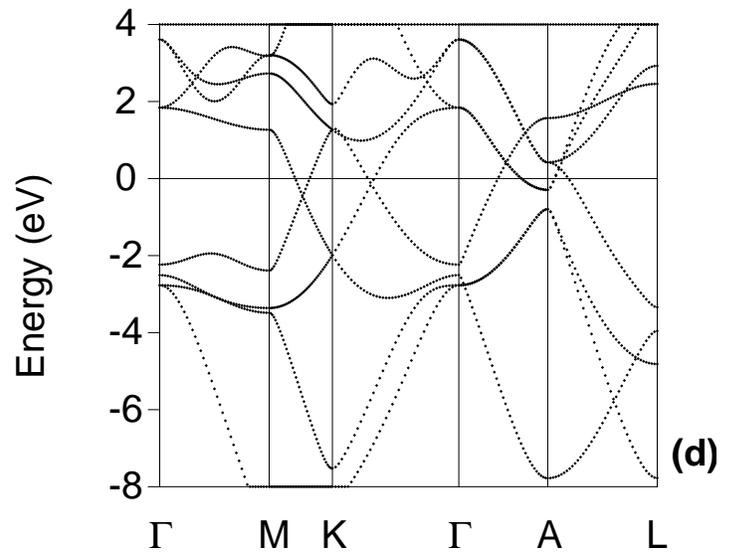
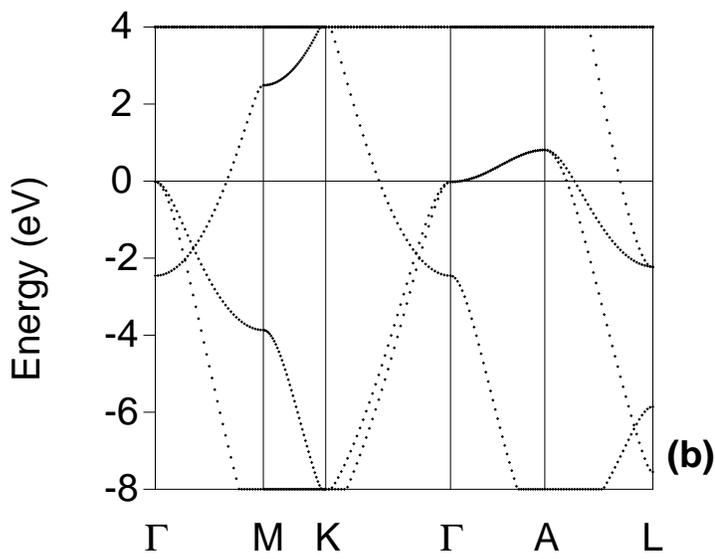
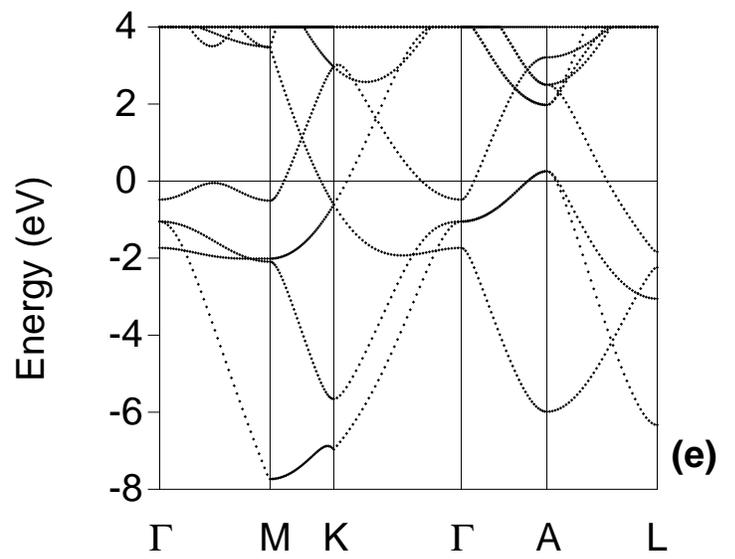
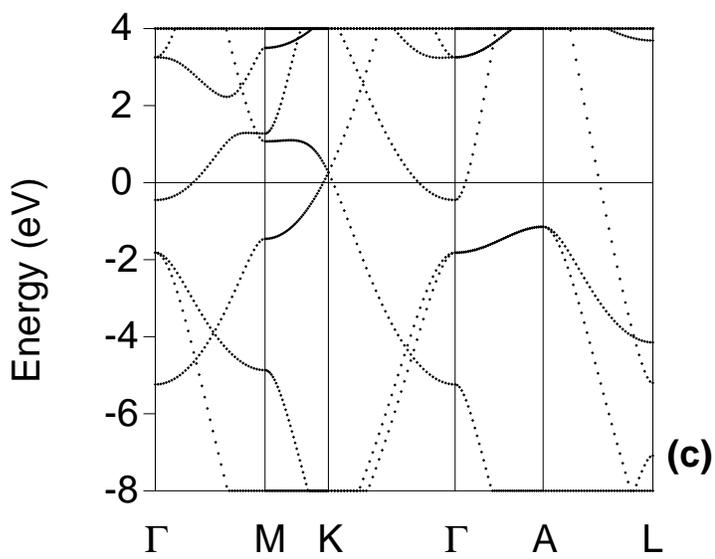
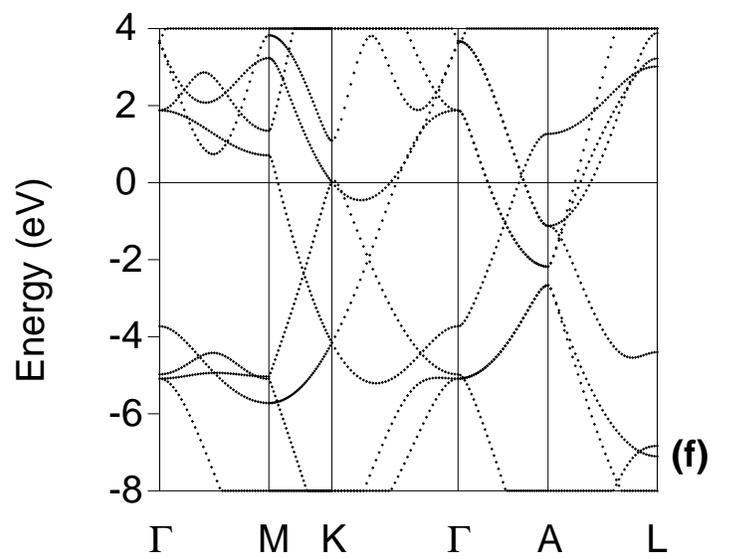
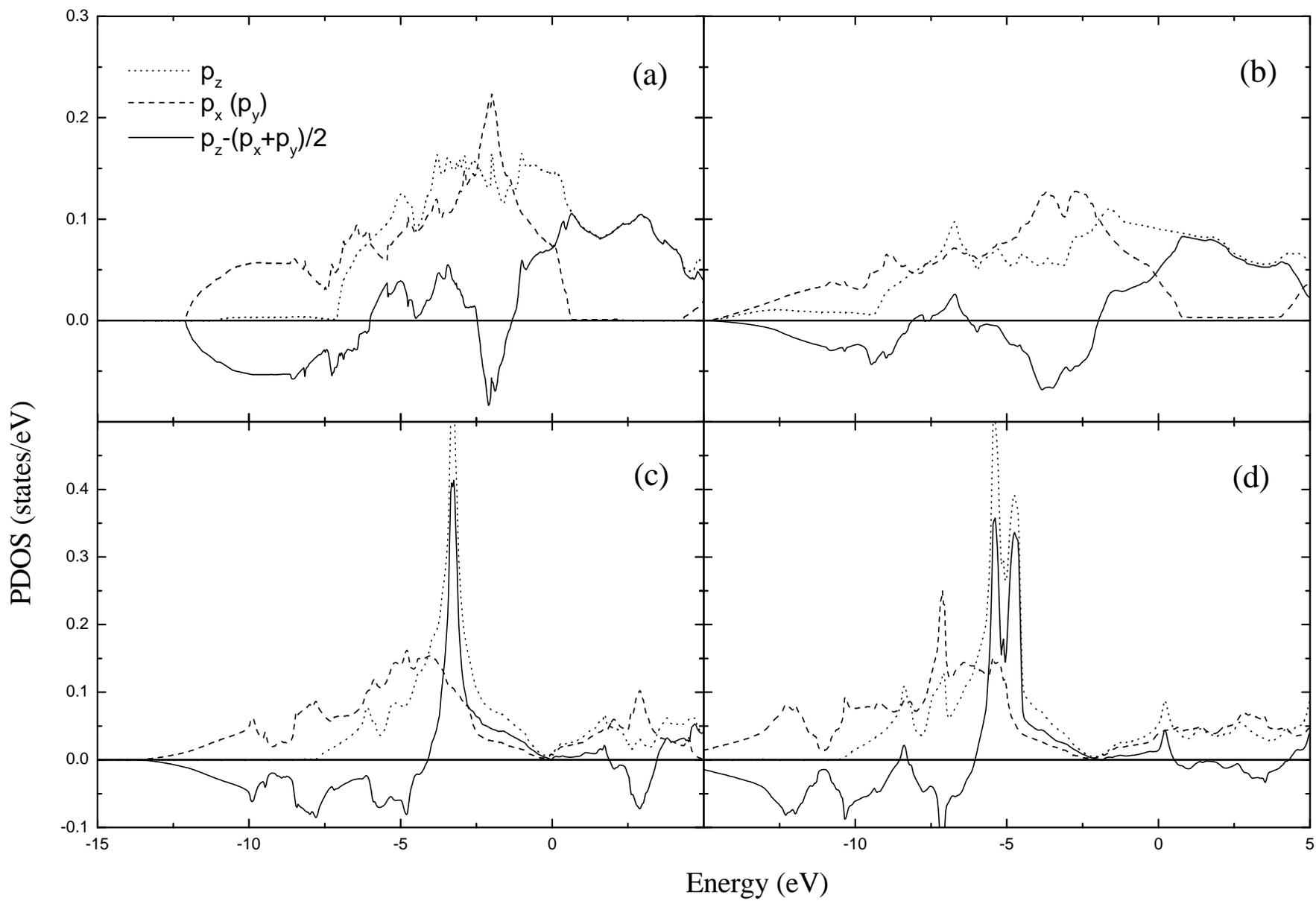
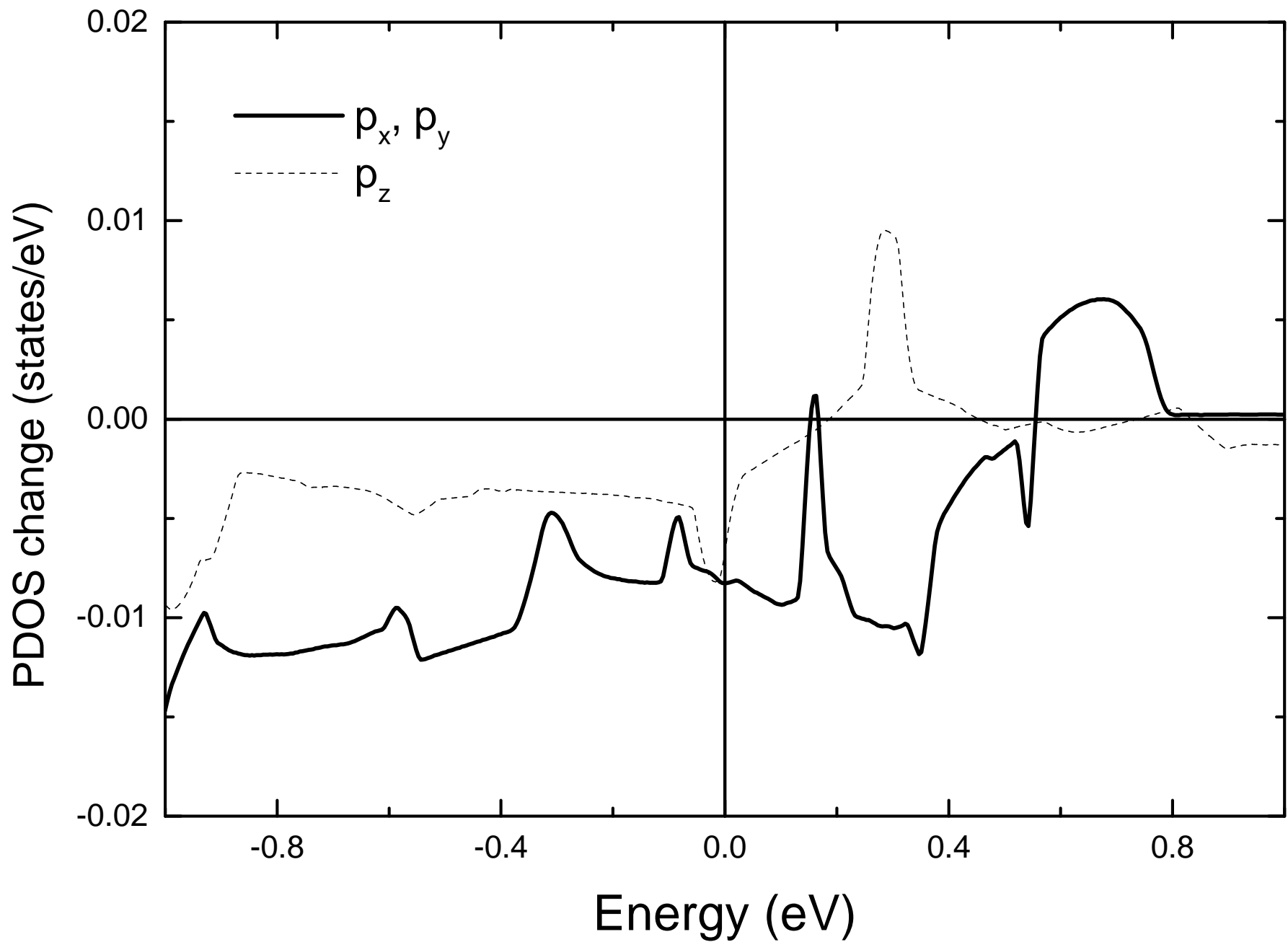
\end{document}